\documentstyle[12pt]{article}

\input{psfig}

\def\Journal#1#2#3#4{{#1} {\bf #2}, #3 (#4)}

\def\PRL{\em Phys. Rev. Lett.}
\def\PRD{{\em Phys. Rev.} D}

\def\be{\begin{equation}}
\def\ee{\end{equation}}
\def\bea{\begin{eqnarray}}
\def\eea{\end{eqnarray}}

\begin{document}







\begin{titlepage}
\begin{center}

\hfill    FERMILAB-CONF-97/089-T\\
\hfill    hep-ph/9704289 \\
\hfill    April, 1997

\vskip .5in

{ \Large \bf Supersymmetric Lepton Flavor Violation\\ at the NLC }

\vskip .3in

{\large Hsin-Chia Cheng}

\vskip .3in

{\it Fermi National Accelerator Laboratory\\
     P.O. Box 500\\
     Batavia, IL 60510}

\end{center}

\vskip .3in

\begin{abstract}

Supersymmetric theories generally have new flavor violation sources in
the squark and slepton mass matrices. If significant lepton flavor 
violation exists, selectron and smuon should be nearly degenerate. 
This leads to the phenomenon of slepton oscillations, which is analogous 
to neutrino oscillations, if sleptons are produced at the Next Linear 
Collider. The direct slepton production at the Next Linear Collider 
provides a much more powerful probe of lepton flavor violation than 
the current bounds from rare processes, such as $\mu \to e\gamma$.

\end{abstract}

\vskip .5in

\begin{center}
Talk given at the 1st Symposium on Flavor-Changing Neutral Currents\\
--Present and Future Studies\\
February 19-21, 1997, Santa Monica, California
\end{center}
\end{titlepage}

\renewcommand{\thepage}{\arabic{page}}
\setcounter{page}{1}
\renewcommand{\thefootnote}{\arabic{footnote}}
\setcounter{footnote}{0}

Weak scale supersymmetry (SUSY) may provide a solution to the gauge 
hierarchy problem and is one of the most attractive candidates beyond 
the standard model (SM). If SUSY is discovered, measuring the masses 
and couplings of the
superpartners will become the focus of study. However, in most SUSY
extensions of the SM, mass matrices of the fermions and their scalar
superpartners are not diagonalized in the same basis. New flavor
mixing matrices $W$, analogous to the CKM matrix,
will appear at the gaugino vertices. 
These new mixing matrices may provide important clues to the flavor
structure and therefore should also be an important subject of study.
Rare flavor-changing processes, such as $\mu\to e\gamma$, and neutral
meson mixing already provide important constraints on these mixing
matrices through the virtual effects of superpartners. However, as
will be seen below, if the superpartners are produced in the future
colliders, it is possible to probe these mixings directly and 
much more powerfully at the colliders. The talk is based on work
done with N. Arkani-Hamed, J.L. Feng, and L.J. Hall \cite{ACFH}.

In this study we will concentrate on the mixing in the slepton
sector. The choice is motivated by the theoretical prejudice that
sleptons are probably lighter than squarks and hence more likely
to be found first, and also by the absence of the lepton flavor
violation in SM, which means that any lepton flavor violation should 
come from the supersymmetric origin. We also specialize to the case
of two generation mixing, in particular $\tilde{e}_R-\tilde{\mu}_R$
mixing for simplicity.

The current experimental bound\cite{Bolton},
$B(\mu\to e\gamma )< 4.9\times 10^{-11}$, put strong
constraints on the slepton masses and mixings. If the mixing
angle $W_{12}=\sin \theta$ is not small, the first two generation
slepton masses have to be quite degenerate for the flavor-changing 
process $\mu\to e\gamma$ to be super-GIM suppressed. The bound
constrains the product $\sin 2\theta \Delta m^2$, where
$\Delta m^2= m^2_{\tilde{e}_R}-m^2_{\tilde{\mu}_R}$, and is plotted
in Fig.~1 for some choices of the SUSY parameters.

If sleptons are produced at colliders, flavor mixing will lead to the
interesting slepton oscillation phenomenon analogous to neutrino 
oscillation. The flavor-violating signal we hope to observe at the 
$e^+ e^-$ colliders is $e^+ e^- \to$ slepton pairs $\to e^{\pm}
\mu^{\mp} \tilde{\chi}^0 \tilde{\chi}^0$. For simplicity, we 
consider the case where the right-handed sleptons decay directly 
to the LSP. For large $\Delta m^2(\gg m\Gamma$, where $\Gamma$ is
the slepton decay width), the $\tilde{e}$ and $\tilde{\mu}$ are
already out of phase when they decay and there is no interference
between them. The flavor-violating cross section is given simply by
multiplying the flavor conserving cross sections by appropriate factors
of the branching ratios $B(\tilde{e}\to \mu \tilde{\chi}^0)
=B(\tilde{\mu}\to e \tilde{\chi}^0)=\frac{1}{2} \sin^2 2\theta$,
When $\Delta m^2 < m\Gamma$, the interference effects become
important. One finds for uncorrelated slepton production,
the flavor-violating cross section will be suppressed by the
factor $\epsilon= (\Delta m^2)^2/[(\Delta m^2)^2 + 4 m^2 \Gamma^2]$.
For correlated production as in the case we study, the formulae 
are somewhat more complicated but contain the similar suppression factor.
The detailed derivation for different cases can be found in 
Ref.[3]. We can see that $\epsilon \simeq 1$ for $\Delta m^2 \gg
m\Gamma$ and $\epsilon \to 0$ as $\Delta m^2 \to 0$ as we expect.
When sleptons are directly produced at high energy colliders, the 
super-GIM suppression of the flavor-violating processes only occurs
when $\Delta m< \Gamma$, in constrast with the low energy 
flavor-violating processes which are suppressed by $\Delta m^2/m^2$.
This makes the high energy colliders a more powerful probe of
the flavor violation than the low energy processes.

Having discussed the flavor-violating cross sections, we consider the
possibility of detecting such flavor-violating signals at future 
colliders. If sleptons are discovered at LEP II, their masses should
be below 85-90 GeV. The constraint from $\mu \to e\gamma$
is strong for such light sleptons. We found that the probing power
at LEP II can still be competitive with
$\mu \to e\gamma$ for moderate $\tan \beta$, though the regions
of parameter space that can be probed are quite limited \cite{ACFH}.
We concentrate on the study of lepton flavor violation at the
Next Linear Collider (NLC) here, with assumed design energy
$\sqrt{s}=500$GeV and luminosity 50 fb$^{-1}/$yr in the $e^+ e^-$
mode. We consider a case with the following SUSY parameters:
$m_{\tilde{e}_R}, m_{\tilde{\mu}_R} \simeq 200\mbox{GeV},\,
M_1=M_2/2=150\mbox{GeV}, \, \mu=-400\mbox{GeV}$.
The LSP is almost pure bino and the cross section has little
dependence on $\mu$ and $M_2$ if they are large. We calculate
the cross section of the flavor-violating signal as a function of
$\sin 2\theta$ and $\Delta m^2$ and the result is shown in Fig.~1.

The major backgrounds come from $W^+ W^-$, $e^{\pm}\nu W^{\mp}$,
$(e^+ e^-) W^+ W^-$, and $\tau^+ \tau^-$ events. Efficient cuts\cite{Becker}
and a right-polarized $e^-$ beam can effectively
isolate the flavor-violating signal. Given a year's running at design
luminosity, the required 5$\sigma$ signal is 3.8(3.6) fb for 
90\%(95\%) right-handed beam polarization. We can see
that NLC is a powerful probe of the flavor-violating parameter space,
extending to $\sin 2\theta \sim 0.1$ and much beyond the bounds
from $\mu \to e\gamma$.

An intriguing feature of the NLC is its ability to run in $e^- e^-$
mode. Although the luminosity may be degraded somewhat from the $e^+ e^-$
mode, the cross section is generally higher for most regions of the parameter
space and many troublesome backgrounds, for example, $W$ pair production,
are completely eliminated. In addition, both beams may be polarized
and there are essentially no backgrounds for RR beam polarization.
The dominant background is $e^-\nu W^-$, arising from imperfect
beam polarization. This can be the most powerful mode for probing
flavor violation. Even without additional cuts, the mixing angle
may be probed down to $\sin 2\theta \sim 0.04$\cite{ACFH}, far beyond the current
bounds. We find that , if sleptons are kinematically accessible at
the NLC, the $e\mu$ signal will provide either stringent limits on
slepton mixing or the exciting discovery of SUSY lepton flavor
violation. 

\begin{figure}
\centerline{\vbox{\psfig{figure=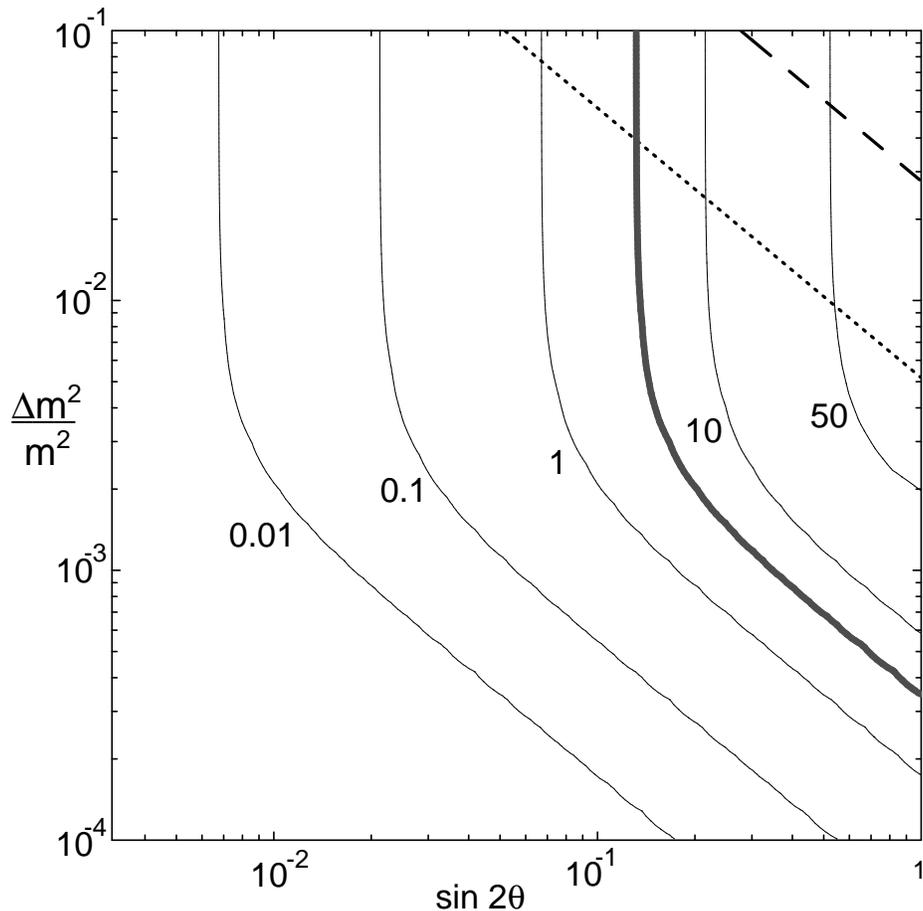,height=5in}}}
\caption{Contours of constant $\sigma(e^+e_R^-\to e^{\pm}\mu^{\mp}
\tilde{\chi}^0 \tilde{\chi}^0)$ (solid) in fb for NLC, with
$m_{\tilde{e}_R}, m_{\tilde{\mu}_R} \simeq 200$GeV,
$M_1=150\mbox{GeV}, \; \mu=-400\mbox{GeV}, \; M_2=300\mbox{GeV}$, and 
$\tan \beta=2$. The thick gray contour represents the experimental reach for
one year's integrated luminosity. The dashed (dotted) contour is the 
current experimental bound from $B(\mu \to e\gamma)=4.9\times 10^{-11}$
for the same parameters and $\tan\beta=2 (50)$, 
$m_{\tilde{e}_L}, m_{\tilde{\mu}_L}$ = 350GeV.
(The cross section contours are basically unchanged for $\tan \beta=50$.) 
\mbox{\hspace{4.5cm}}
}
\end{figure}



\end{document}